\documentclass[twocolumn]{aastex7}

\usepackage{amsmath}
\usepackage{multirow}
\usepackage{threeparttable}

\def\ion#1#2{#1$\;${\footnotesize\rm{#2}}\relax}

\begin{document}

\title{Optically Overluminous TDEs: Outflow Properties and Implications for Extremely Relativistic Disruptions}

 
\author[0000-0001-6747-8509]{Yuhan Yao}
\email[show]{yuhanyao@berkeley.edu}
\affiliation{Miller Institute for Basic Research in Science, 206B Stanley Hall, Berkeley, CA 94720, USA}
\affiliation{Department of Astronomy, University of California, Berkeley, CA 94720-3411, USA}

\author[0000-0002-8297-2473]{Kate D. Alexander}\email{kdalexander@arizona.edu}
\affiliation{Department of Astronomy/Steward Observatory, 933 North Cherry Avenue, Room N204, Tucson, AZ 85721-0065, USA}

\author[0000-0002-1568-7461]{Wenbin Lu}\email{wenbinlu@berkeley.edu}
\affiliation{Department of Astronomy, University of California, Berkeley, CA 94720-3411, USA}
\affiliation{Theoretical Astrophysics Center, University of California, Berkeley, CA 94720, USA}

\author[0000-0001-8426-5732]{Jean J. Somalwar}\email{jsomalwa@caltech.edu}
\affil{Cahill Center for Astronomy and Astrophysics, California Institute of Technology, MC 249-17, 1200 E California Boulevard, Pasadena, CA 91125, USA}

\author[0000-0002-7252-5485]{Vikram Ravi}\email{v.vikram.ravi@gmail.com}
\affil{Cahill Center for Astronomy and Astrophysics, California Institute of Technology, MC 249-17, 1200 E California Boulevard, Pasadena, CA 91125, USA}

\author[0000-0002-7706-5668]{Ryan Chornock}\email{chornock@berkeley.edu}
\affiliation{Department of Astronomy, University of California, Berkeley, CA 94720-3411, USA}

\author[0000-0003-4768-7586]{Raffaella Margutti}\email{rmargutti@berkeley.edu}
\affiliation{Department of Astronomy, University of California, Berkeley, CA 94720-3411, USA}
\affiliation{Department of Physics, University of California, 366 Physics North MC 7300, Berkeley, CA 94720, USA}

\author[0000-0001-8472-1996]{Daniel A.~Perley}\email{D.A.Perley@ljmu.ac.uk}
\affiliation{Astrophysics Research Institute, Liverpool John Moores University, IC2, Liverpool Science Park, 146 Brownlow Hill, Liverpool L3 5RF, UK}

\author[0000-0003-3124-2814]{James C. A. Miller-Jones}\email{james.miller-jones@curtin.edu.au}
\affiliation{International Centre for Radio Astronomy Research Curtin University, GPO Box U1987, Perth, WA 6845, Australia}

\author[0000-0001-7833-1043]{Paz Beniamini}\email{pazb@openu.ac.il}
\affiliation{Department of Natural Sciences, The Open University of Israel, P.O Box 808, Ra’anana 4353701, Israel}
\affiliation{Astrophysics Research Center of the Open university (ARCO), The Open University of Israel, P.O Box 808, Ra’anana 4353701, Israel}


\author[0000-0002-8070-5400]{Nayana A. J.}\email{nayana@berkeley.edu}
\affiliation{Department of Astronomy, University of California, Berkeley, CA 94720-3411, USA}

\author[0000-0002-7777-216X]{Joshua S. Bloom}\email{joshbloom@berkeley.edu}
\affiliation{Department of Astronomy, University of California, Berkeley, CA 94720-3411, USA}
\affiliation{Lawrence Berkeley National Laboratory, 1 Cyclotron Road, MS 50B-4206, Berkeley, CA 94720, USA}

\author[0000-0003-0528-202X]{Collin T. Christy}\email{collinchristy@arizona.edu}
\affiliation{Department of Astronomy/Steward Observatory, 933 North Cherry Avenue, Room N204, Tucson, AZ 85721-0065, USA}

\author[0000-0002-3168-0139]{Matthew J. Graham}\email{mjg@caltech.edu}
\affiliation{Cahill Center for Astronomy and Astrophysics, California Institute of Technology, MC 249-17, 1200 E California Boulevard, Pasadena, CA 91125, USA}

\author[0000-0001-5668-3507]{Steven L. Groom}\email{sgroom@ipac.caltech.edu}
\affiliation{IPAC, California Institute of Technology, 1200 E. California Blvd, Pasadena, CA 91125, USA}

\author[0000-0002-5698-8703]{Erica Hammerstein}\email{ekhammer@berkeley.edu}
\affiliation{Department of Astronomy, University of California, Berkeley, CA 94720-3411, USA}

\author[0000-0003-3367-3415]{George Helou}\email{gxh@ipac.caltech.edu}
\affiliation{IPAC, California Institute of Technology, 1200 E. California Blvd, Pasadena, CA 91125, USA}

\author[0000-0002-5619-4938]{Mansi M. Kasliwal}\email{mansi@astro.caltech.edu}
\affiliation{Cahill Center for Astronomy and Astrophysics, California Institute of Technology, MC 249-17, 1200 E California Boulevard, Pasadena, CA 91125, USA}

\author[0000-0001-5390-8563]{S.~R.~Kulkarni}\email{srk@astro.caltech.edu}
\affiliation{Cahill Center for Astronomy and Astrophysics, California Institute of Technology, MC 249-17, 1200 E California Boulevard, Pasadena, CA 91125, USA}

\author[0000-0003-2451-5482]{Russ R. Laher}\email{laher@ipac.caltech.edu}
\affiliation{IPAC, California Institute of Technology, 1200 E. California Blvd, Pasadena, CA 91125, USA}

\author[0000-0003-2242-0244]{Ashish~A.~Mahabal}\email{aam@astro.caltech.edu}
\affiliation{Cahill Center for Astronomy and Astrophysics, California Institute of Technology, MC 249-17, 1200 E California Boulevard, Pasadena, CA 91125, USA}
\affiliation{Center for Data Driven Discovery, California Institute of Technology, Pasadena, CA 91125, USA}

\author[0000-0002-6966-5946]{Jérémy Neveu}\email{jeremy.neveu@ijclab.in2p3.fr}
\affiliation{Universit\'e Paris-Saclay, CNRS, IJCLab, 91405, Orsay, France}

\author[0000-0002-0387-370X]{Reed Riddle}\email{riddle@caltech.edu}
\affiliation{Caltech Optical Observatories, California Institute of Technology, Pasadena, CA 91125, USA}

\author[0000-0001-7062-9726]{Roger Smith}\email{rsmith@astro.caltech.edu}
\affiliation{Caltech Optical Observatories, California Institute of Technology, Pasadena, CA  91125, USA}

\author[0000-0002-3859-8074]{Sjoert van Velzen}\email{sjoert@strw.leidenuniv.nl}
\affiliation{Leiden Observatory, Leiden University, Postbus 9513, 2300 RA, Leiden, The Netherlands}

\begin{abstract}
Recent studies suggest that tidal disruption events (TDEs) with off-axis jets may manifest as optically overluminous events. 
To search for jet signatures at late times, we conducted radio observations of eight such optically overluminous ($M_{g, \rm peak} < -20.8$\,mag) TDEs with the Very Large Array. 
We detect radio counterparts in four events. 
The observed radio luminosities ($L_{\rm 6\,GHz} \sim 10^{38}$–-$10^{39}$\,erg\,s$^{-1}$) are two orders of magnitude lower than those of on-axis jetted TDEs, and we find no evidence for off-axis jets within rest-frame time of 3 yrs.
Two of them (AT2022hvp and AT2021aeou) exhibit evolving radio emission, consistent with synchrotron emission from non-relativistic outflows launched near the time of first optical light. 
Two events (AT2020ysg and AT2020qhs) show no statistically significant variability, which can be attributed to either non-relativistic outflows or pre-existing active galactic nuclei.  
Compared to a control sample of fainter TDEs with $M_{g, \rm peak} > -20.5$\,mag observed at similar rest-frame timescales ($t_{\rm rest} \sim 1.5$\,yr), our sample shows systematically more luminous radio emission, suggesting that optically overluminous TDEs may launch more powerful prompt non-relativistic outflows. We speculate that strong general relativistic effects near high-mass black holes ($M_{\rm BH} \sim 10^8\,M_\odot$) may play a key role. 
These findings motivate further investigation into the nature of relativistic disruptions around massive black holes and the physical conditions necessary for jet formation.
\end{abstract}


\section{Introduction} 

Tidal disruption events (TDEs) are rare electromagnetic transients where a star is torn apart by the tidal forces of a massive black hole. In some cases, such events are accompanied by the ejection of material in the form of collimated relativistic jets, which emit prompt X-ray light and lower frequency afterglows (see \citealt{DeColle2020, Alexander2020} for reviews). So far, only four on-axis jetted TDEs have been identified, including Sw\,J1644+57\footnote{We note that an alternative interpretation is that Sw\,J1644+57 harbors a slightly off-axis jet \citep{Beniamini2023}.} \citep{Bloom2011, Burrows2011, Levan2011, Zauderer2011}, Sw\,J2058+05 \citep{Cenko2012, Pasham2015}, Sw\,J1112-82 \citep{Brown2015, Brown2017_J1112}, and AT2022cmc \citep{Andreoni2022, Pasham2023}. The volumetric rate of on-axis jetted TDEs is found to be 0.01--0.07\,Gpc$^{-3}$\,yr$^{-1}$ \citep{Sun2015, Andreoni2022}. Assuming a relativistic beaming factor of $f_{\rm b}\approx 0.01$, the intrinsic rate of jetted TDEs\footnote{Strictly speaking, this refers to TDEs with prompt relativistic jets. In this paper, we do not consider the possibility of relativistic jets launched significantly after disruption.} (1--7\,Gpc$^{-3}$\,yr$^{-1}$) appears to be a tiny fraction ($\lesssim\!0.1$--1\%) of the the total TDE rate of $\sim\!10^3$\,Gpc$^{-3}$\,yr$^{-1}$ \citep{Sazonov2021, Yao2023, Masterson2024}. It is possible that misaligned precessing jets can be choked by the accretion disk wind \citep{teboul23_choked_jets, Lu2024_choked_jet}. 

In the UV and optical band, the TDE population spans a wide range of peak luminosities and spectral subtypes \citep{vanVelzen2020}. Based on the existence of broad (full-width at half-maximum of $> 5\times 10^3\,{\rm km\,s^{-1}}$) emission lines, \citet{vanVelzen2021} proposed a classification scheme dividing TDEs into three subclasses: TDE-H, TDE-He, and TDE-H+He. 
More recently, \citet{Hammerstein2023} identified four TDEs with high luminosities and featureless optical spectra, and additional similar events have been reported by \citet{Yao2023}. While many of these overluminous TDEs lack spectral features, some do exhibit broad lines (e.g., see \citealt{Kumar2024} and \S\ref{sec:sample}). Conversely, there are also examples of optically subluminous TDEs with featureless spectra \citep[e.g.,][]{Yao2022_21ehb}.

\begin{figure}[htbp!]
    \centering
    \includegraphics[width=\columnwidth]{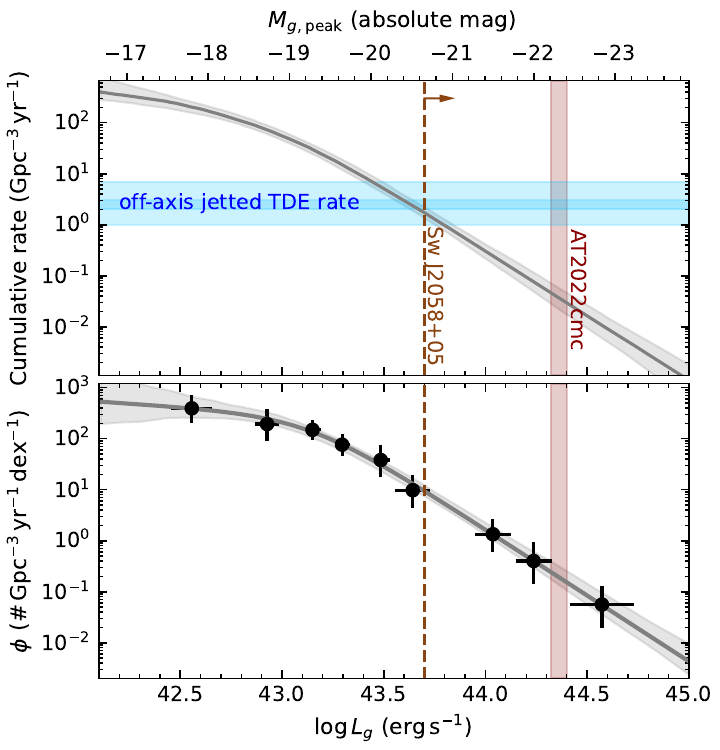}
    \caption{Cumulative optical TDE rate (upper panel) and rest-frame $g$-band ($\nu_{\rm rest}=6.3\times10^{14}$\,Hz) luminosity function (bottom panel; \citealt{Yao2023}). 
    The vertical lines mark the the peak $L_g$ of two on-axis jetted TDEs \citep{Pasham2015, Andreoni2022, Yao2024}. \label{fig:Lg_func_rate}
    }
\end{figure}

\citet{Andreoni2022} suggested the possibility that the population of off-axis jetted TDEs might manifest themselves as slowly evolving, overluminous, featureless blue nuclear transients in optical sky surveys. 
This hypothesis is based on the fact that, among known on-axis jetted TDEs, the peak (rest-frame) UV and optical spectral energy distribution (SED) has been observed in Sw\,J2058+05 \citep{Pasham2015} and AT2022cmc \citep{Andreoni2022, Yao2024, Hammerstein2025}, and that both events exhibit thermal SEDs that can be described by a blackbody with low values of $M_{g, \rm peak}$ (see Figure~\ref{fig:Lg_func_rate}) and featureless spectra.
Further evidence comes from the volumetric rate of optically overluminous TDEs, which is ${\rm few} \times $\,Gpc$^{-3}$\,yr$^{-1}$ (see Figure~\ref{fig:Lg_func_rate}) --- a rate consistent with expectations for off-axis jetted TDEs.

In this work, we test this hypothesis using late-time radio observations of eight optically overluminous TDEs discovered by the Zwicky Transient Facility (ZTF; \citealt{Bellm2019b, Graham2019, Masci2019, Dekany2020}). 
In the off-axis jet scenario, radio emission is expected to peak on a timescale that depends on viewing angle 
and eventually resemble the light curve of an on-axis jet \citep{Ryan2020, Beniamini2020}. Therefore, if most optically overluminous TDEs harbor off-axis relativistic jets, their late-time radio luminosities should be comparable to those of known jetted TDEs.

UT time is used throughout the paper. We adopt a standard $\Lambda$CDM cosmology with $\Omega_{\rm M} = 0.3$, $\Omega_{\Lambda}=0.7$, and $H_0=70\,{\rm km\,s^{-1}\,Mpc^{-1}}$.
Uncertainties are reported at the 68\% confidence intervals unless otherwise noted, and upper limits are reported at 3$\sigma$.

\section{Sample Selection and Observations}\label{sec:sample}

We collected all TDEs that were first detected by ZTF from 2019 January 1 to 2022 May 1, resulting in 62 events. The TDEs were photometrically selected using a custom filter \citep{vanVelzen2021} built upon the \texttt{AMPEL} broker \citep{Nordin2019}, and spectroscopically classified. 
We modeled their UV--optical light curves and host galaxies using the procedures outlined in \citet{Yao2023}. 
The best-fit light curve model provides peak rest-frame $g$-band absolute magnitude $M_{g, \rm peak}$\footnote{For light curves with multiple peaks, $M_{g, \rm peak}$ is measured for the most luminous peak.} and the functional form (power-law or Gaussian) that best describes the rising part of the light curves. For a power-law rise, the first-light epoch $t_{\rm fl}$ is given by the best-fit model. For a Gaussian rise, we define $t_{\rm fl} \equiv t_{\rm peak} - 3\sigma_{\rm rise}$, where $t_{\rm peak}$ is time of the optical peak and $\sigma_{\rm rise}$ is the Gaussian rise time. 
Hereafter, we use $t_{\rm rest}$ to denote rest-frame time with respect to $t_{\rm fl}$.

\begin{figure}[htbp!]
    \centering
    \includegraphics[width=\columnwidth]{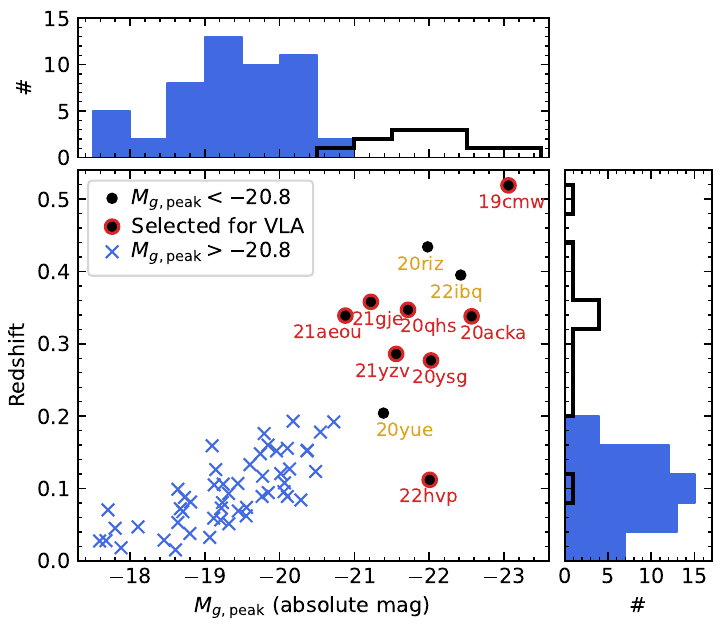}
    \caption{ZTF-selected TDEs (2019.0--2022.3) on the diagram of redshift vs. peak rest-frame $g$-band absolute magnitude. 
    Overluminous TDEs selected for VLA observations are annotated. \label{fig:z_Mgpeak}}
\end{figure}

\begin{figure}[htbp!]
     \centering
     \includegraphics[width = \columnwidth]{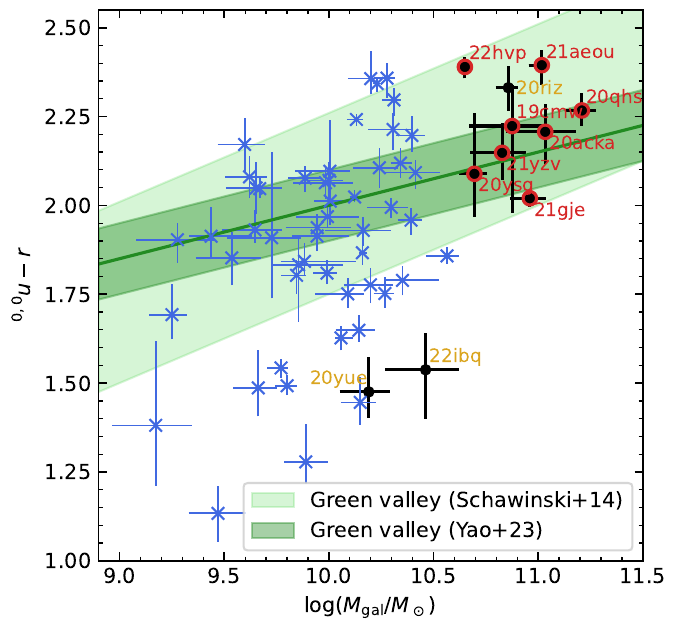}
     \caption{The same sample as in Figure~\ref{fig:z_Mgpeak}, but in the galaxy color-mass diagram. 
     TDEs selected for VLA observations are hosted by higher-mass galaxies. 
     \label{fig:Mgal_umr} }
 \end{figure}

Figure~\ref{fig:z_Mgpeak} shows the distribution of 62 TDEs on the panel of redshift vs. $M_{g, \rm peak}$. Figure~\ref{fig:Mgal_umr} shows the host galaxy total stellar mass $M_{\rm gal}$ versus Galactic extinction-corrected, synthetic rest-frame $u-r$ color. A total of 11 TDEs have $M_{g, \rm peak}<-20.8$\,mag. 
The threshold of $-20.8$\,mag is a somewhat arbitrary cut to separate overluminous and normal-luminosity events, as the luminosity function is a continuous power-law in this range (see Figure~\ref{fig:Lg_func_rate}). 
As can be seen, due to the small intrinsic rate of overluminous TDEs, they are generally selected at much higher redshifts ($z>0.1$) and were mostly missed by previous radio TDE follow up programs. We selected 8 objects spanning a range of $M_{g,\rm peak}$ with different spectroscopic and light curve shapes for radio observations. 

\begin{figure*}[htbp!]
     \centering
     \includegraphics[width = \textwidth]{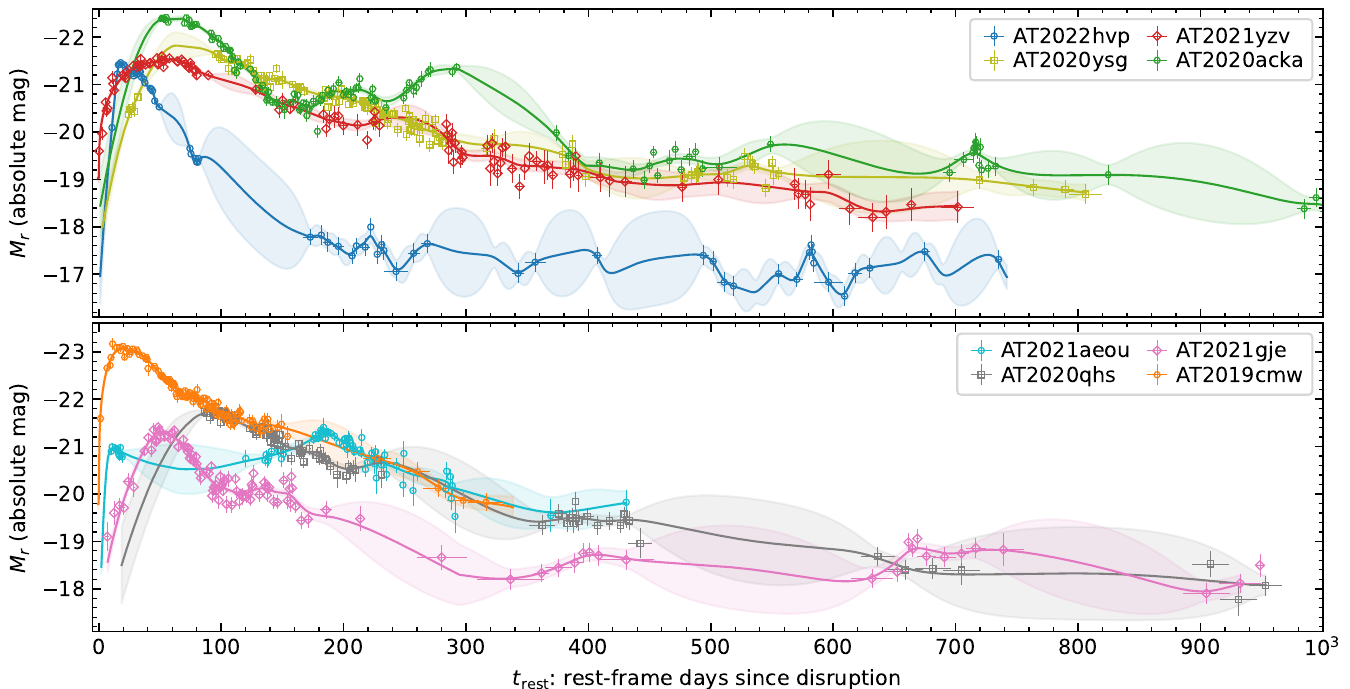}
     \caption{ZTF $r$-band light curves of our sample.
     To guide the eye in tracking the light curve evolution, we fit the data using a combination of functional forms and Gaussian process smoothing, following procedures described in Appendix~B.4 of \citet{Yao2020}. The best-fit models are shown as solid lines, with the 68\% confidence intervals displayed in transparent colors.
     Absolute magnitude is computed using $ M = m - 5\,{\rm log_{10}}[D_L / (\rm 10\,pc) ] + 2.5 \,{\rm log_{10}} (1+z) $, where the last term is a rough estimation of the $K$-correction. 
     \label{fig:lumTDE_lc} }
 \end{figure*}

Basic properties of the sample are summarized in Table~\ref{tab:sample}. Our sample spans a range of $M_{g, \rm peak}$ and various spectral subtypes. Figure~\ref{fig:lumTDE_lc} shows their ZTF $r$-band light curves obtained through forced photometry \citep{Masci2023}.  As can be seen, most of them exhibit late-time plateau that has been uniquely observed in TDEs \citep{Mummery2024_fundamental_scaling_relation}.

We adopt the spectroscopic classification nomenclature introduced in \citet{vanVelzen2021,Hammerstein2023,Yao2023}, where optical TDEs are divided into five subtypes: TDE-H, TDE-He, TDE-H+He, TDE-featureless, and TDE-coronal. 
In our sample, five objects (AT2020ysg, AT2021yzv, AT2020acka, AT2020qhs, AT2019cmw) have been previously reported as TDEs with assigned subtypes in refereed articles \citep{Hammerstein2023, Yao2023}, two objects (AT2022hvp, AT2021gje) have been reported to the transient name server (TNS) as TDEs \citep{Fulton2022_22hvp, Hammerstein2021_21gje_CR}, and AT2021aeou has not been classified before. 

\begin{deluxetable*}{lllllcccc}[htbp!]
	\tablecaption{Basic Information of 8 optically overluminous TDEs selected for VLA observations.\label{tab:sample}}
 \tablehead{
    \colhead{IAU name}
    & \colhead{ZTF name}
    & \colhead{Redshift}
    & \colhead{TDE Report}
    & \colhead{Spectral Subtype}
    & \colhead{$M_{g,\rm peak}$}
    & \colhead{$t_{\rm fl}$ MJD}
    & \colhead{$\sigma_\ast$ (km\,s$^{-1}$)\tablenotemark{c}}
    & \colhead{$M_{\rm BH}$\tablenotemark{c}}
    }
    \startdata
    AT2022hvp & ZTF22aagyuao & 0.112 & \citet{Fulton2022_22hvp} & TDE-He\tablenotemark{a}
    & $-22.01$ & 59676.3 &  $134.96 \pm 11.03$ & $7.74 \pm 0.34$\\
    AT2020ysg & ZTF20abnorit & 0.277 & \citet{Hammerstein2023} & TDE-He\tablenotemark{a}
    & $-22.02$ & 59015.3 & $157.78 \pm 13.03$ & $8.04 \pm 0.33$\\
    AT2021yzv & ZTF21abxngcz & 0.286 & \citet{Yao2023} & TDE-featureless & 
    $-21.56$ & 59435.3 & $146.38 \pm 20.78$ & $7.90 \pm 0.40$\\ 
    AT2020acka & ZTF20acwytxn & 0.338 & \citet{Yao2023} & TDE-featureless & 
    $-22.57$ & 59125.8 & $174.47 \pm 25.30$ & $8.23 \pm 0.40$\\
    AT2021aeou & ZTF21abvpudz & 0.339 & This paper & TDE-featureless\tablenotemark{a}
    & $-20.88$ & 59439.5 & & $8.30 \pm 0.83$\\
    AT2020qhs  & ZTF20abowque & 0.347 & \citet{Hammerstein2023} & TDE-featureless 
    & $-21.72$ & 58945.1 & $188.69 \pm 37.86$ & $8.38 \pm 0.48$\\
    AT2021gje  & ZTF21aapvvtb & 0.358 & \citet{Hammerstein2021_21gje_CR} & TDE-He-pec\tablenotemark{b} & $-21.22$  & 59240.5 & $132.00 \pm 8.55$ & $7.70 \pm 0.32$\\
    AT2019cmw  & ZTF19aaniqrr & 0.519 & \citet{Yao2023}  & TDE-featureless & $-23.06$ & 58558.3 & & $8.07 \pm 0.87$\\
 \enddata
 \tablenotetext{a}{See optical spectra presented in Appendix~\ref{sec:opt_spec}.}
 \tablenotetext{b}{AT2021gje exhibits peculiar optical spectral properties. Detailed analysis will be presented in W. Wu et al. (in prep).}
 \tablenotetext{c}{See Appendix~\ref{sec:velocity_dispersion} for the methods to estimate velocity dispersion $\sigma_\ast$ and black hole mass $M_{\rm BH}$.}
\end{deluxetable*}

The detailed single-object papers (including optical spectral observations) of AT2019cmw and AT2021gje will be presented by J. Wise et al. (in prep) and W. Wu et al. (in prep), respectively. 
For three objects (AT2022hvp, AT2020ysg, and AT2021aeou), we present optical spectroscopy in Appendix~\ref{sec:opt_spec}, which supports the TDE redshift and subtype classification shown in Table~\ref{tab:sample}.

We obtained radio observations using NSF's Karl G. Jansky Very Large Array (VLA; \citealt{Perley2011}). The search epoch is conducted in C-band under program 23A-280 (PI: Y.~Yao). Follow-up observations for four detected TDEs are obtained through programs 23A-413, and 24A-290 (PI: Y.~Yao). We also include data of AT2022hvp obtained under the VLA large program 20B-377 (PI: K.~D.~Alexander). 
The data were analyzed following the standard radio continuum image analysis procedures in the Common Astronomy Software Applications (\texttt{CASA}; \citealt{CASATeam2022}). 
We used \texttt{tclean} to produce radio images. The flux density was measured as the maximum pixel value within a region the size of the synthesized beam, centered on the optical coordinates of the TDE. The uncertainty was estimated as the root-mean-square (rms) of the pixel values in a nearby source-free region of the image. 
The results are presented in Appendix~\ref{sec:tables} (Table~\ref{tab:vla}).

\section{Non-relativistic Outflow modeling}

\begin{figure*}
    \centering
    \includegraphics[width=0.82\textwidth]{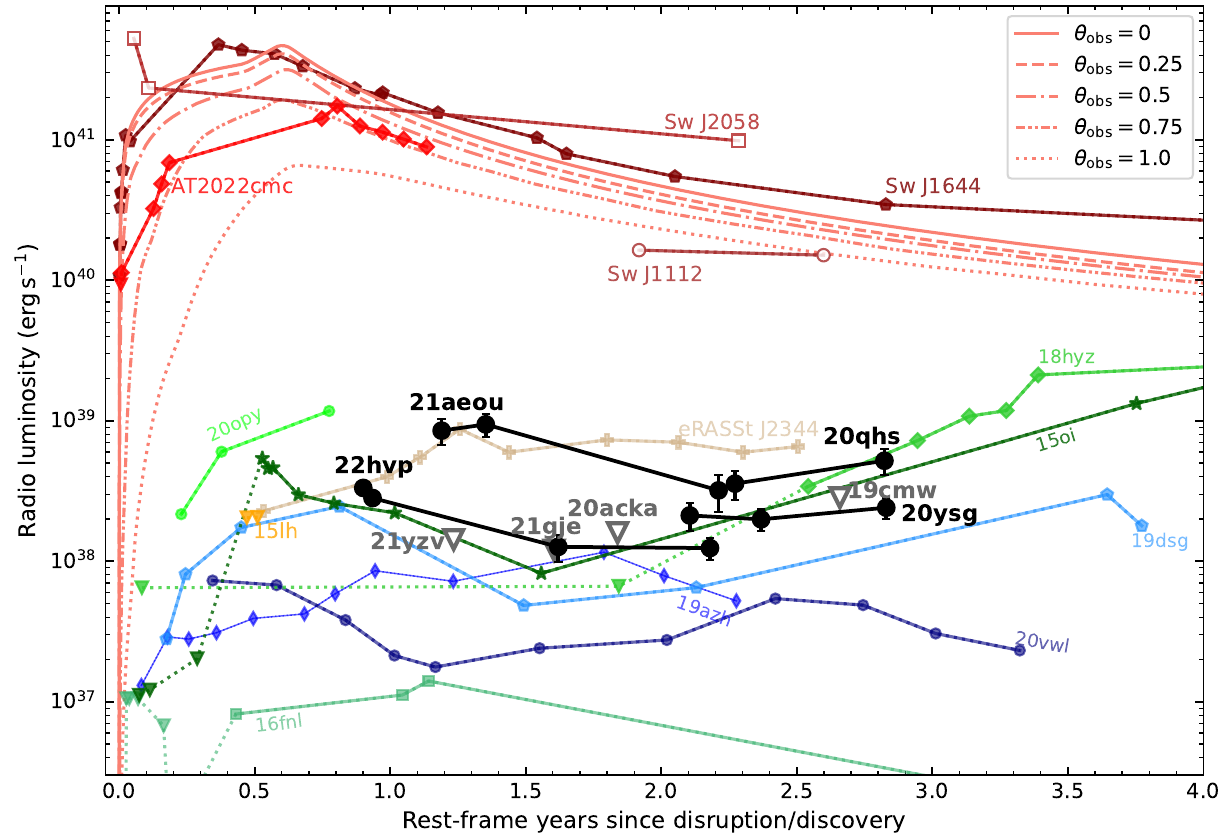}
    \caption{The black data points and gray downward triangles show VLA detections and $3\sigma$ upper limits of optically overluminous TDEs.
    In comparison, we show radio ($\sim6$\,GHz) light curves of on-axis jetted TDEs (red lines) and other TDEs with high-cadence radio observations in the literature.
    We also plot 6\,GHz light curves from a jet with the same properties as found for the best-fit model of Sw\,J1644 \citep{Beniamini2023}, but at different $\theta_{\rm obs}$ (see \S\ref{subsec:no_jets}). 
    References: ASASSN-15oi \citep{Horesh2021, Hajela2025}, iPTF16fnl \citep{Alexander2020, Horesh2021, Cendes2024}, AT2018hyz \citep{Cendes2022_18hyz, Cendes2024}, 
    AT2019azh \citep{Goodwin2022_19azh}, AT2019dsg \citep{Cendes2021_19dsg, Cendes2024}, AT2020opy \citep{Goodwin2023_20opy}, AT2020vwl \citep{Goodwin2023_20vwl, Goodwin2025_20vwl}, eRASSt J234402.9-52640 (also AT2020wjw) \citep{Goodwin2024}; Sw\,J1644+57 \citep{Zauderer2011, Zauderer2013, Eftekhari2018}, Sw\,J2058+05 \citep{Cenko2012, Pasham2015, Brown2017_J1112}, Sw\,J1112-82 \citep{Brown2017_J1112}, and AT2022cmc \citep{Andreoni2022, Rhodes2025}. 
    We also show two overluminous TDE candidates with radio observations reported in the literature: ASASSN-15lh ($M_{g, \rm peak}\sim -23.4$\,mag; \citealt{Kool2015, Leloudas2016, Margutti2017}) and eRASSt\,J2344 ($M_{g, \rm peak}\sim -21.8$\,mag; \citealt{Homan2023, Goodwin2024}).
    \label{fig:radio_lc}}
\end{figure*}

Figure~\ref{fig:radio_lc} presents the 6\,GHz C-band light curves and upper limits for our sample, alongside a comparison with well-studied TDEs from the literature. Four objects (AT2021yzv, AT2020acka, AT2021gje, and AT2019cmw) are not detected. 
We perform spectral fitting to assess whether or not the four events with detected radio emission are consistent with synchrotron self absorption expected in a newly launched non-relativistic TDE outflow. 

We assume the electrons in the shock are accelerated into a power-law distribution, $N(\gamma_{\rm e}) \propto \gamma_{\rm e}^{-p}$ for $\gamma_{\rm e} \geq \gamma_{\rm m}$, where $\gamma_{\rm m}$ is the minimum Lorentz factor of the relativistic electrons. 
A fraction $\epsilon_{\rm e}$ of the shock energy goes into relativistic electrons, and a fraction $\epsilon_{\rm B}$ of shock energy goes into magnetic energy density.
The critical electron Lorentz factor at which synchrotron cooling time equals to the dynamical time is $\gamma_{\rm c}$. The characteristic synchrotron frequencies for electrons with $\gamma_{\rm e} =\gamma_{\rm m}$ and $\gamma_{\rm e} =\gamma_{\rm c}$ are denoted as $\nu_{\rm m}$ and $\nu_{\rm c}$, respectively. The self-absorption frequency is denoted as $\nu_{\rm a}$, below which the system is optically thick to its own synchrotron emission. 

We assume that $\nu_{\rm p}$ is associated with $\nu_{\rm a}$ (i.e., $\nu_{\rm m} \ll \nu_{\rm a} =\nu_{\rm p} \ll \nu_{\rm c}$), which is generally the case for a non-relativistic outflow. 
The radio SED follows a smoothed power-law:
\begin{align}
    L_\nu = L_{\nu,\rm p} \left[ \left( \frac{\nu}{\nu_{\rm p}}\right)^{-s \beta_1} + \left( \frac{\nu}{\nu_{\rm p}}\right)^{-s \beta_2} \right]^{-1/s} \label{eq:GSbrokenPL}
\end{align}
where $\nu$ and $L_\nu$ are quantities in the object's rest-frame, $\beta_1=5/2$ and $\beta_2=-(p-1)/2$ are the asymptotic spectral indices below and above the break, and $s=1.25-0.18p$ is a smoothing parameter \citep{Granot2002}.

\subsection{AT2020hvp and AT2021aeou}

\begin{figure}[htbp]
    \includegraphics[width=0.55\columnwidth]{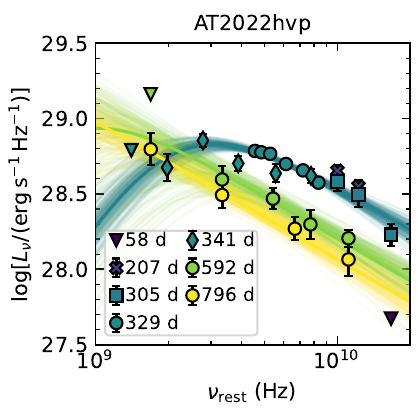}
    \includegraphics[width=0.435\columnwidth]{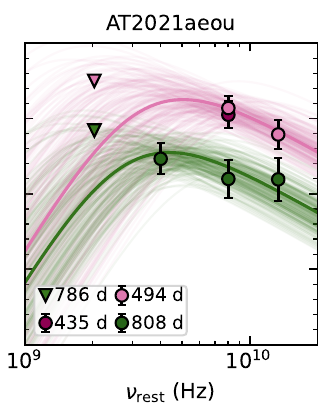}
    \caption{Radio SEDs of AT2022hvp and AT2021aeou, overplotted with the best-fit broken power-law models. \label{fig:radio_sed_22hvp_21aeou}}
\end{figure}

The SED evolution of AT2022hvp and AT2021aeou is shown in Figure~\ref{fig:radio_sed_22hvp_21aeou}. The the peak specific luminosity $L_{\nu,\rm p}$ and the peak frequency $\nu_{\rm p}$ can be derived for each epoch. 
We perform the fit using the Markov chain Monte Carlo (MCMC) approach with \texttt{emcee} \citep{Foreman-Mackey2013}.
Both detections and non-detections are incorporated in the fitting following the procedures outlined in \citet{Laskar2014} and \citet{Eftekhari2024}. 
For AT2022hvp, we fix the value of $p$ to be the same across different epochs.
Due to the partial SED coverage in AT2021aeou, we fix $p=2.7$ (cf., \citealt{Alexander2016, Cendes2021_19dsg, Goodwin2022_19azh}). 

\begin{table*}[htbp]
    \caption{Radio spectral fit parameters and inferred non-relativistic outflow quantities for AT2022hvp and AT2021aeou.}
    \centering
    \begin{tabular}{cccccccccc}
    \hline\hline
      Name   & $t_{\rm rest}$ & $p$ & log$\nu_{\rm p}$ & log$L_{\nu ,\rm p}$ & log$E_{\rm eq}$ & log$R_{\rm eq}$ & log$n_{\rm e}$ & $\beta$ & log$M_{\rm ej}$\\
      & (d)  & & (Hz) & $({\rm erg\,s^{-1}\,Hz^{-1}})$ & (erg) & (cm) & $(\rm cm^{-3})$ & & ($M_\odot$)\\
    \hline
      \multirow{3}{*}{AT2022hvp} & 329--341 & \multirow{3}{*}{$2.86_{-0.21}^{+0.24}$} & $9.32_{-0.04}^{+0.05}$ & $29.17\pm0.04$ & $50.11_{-0.16}^{+0.17}$ & $17.18_{-0.04}^{+0.05}$ & $1.98\pm0.21$ & $0.17\pm0.01$ & $-2.27\pm0.17$\\
      & 592 & & $<9.50$ & $>28.77$  & $>49.77$ &  $>16.86$ & $<2.72$ & $>0.05$ & $>-2.78$\\
      & 796 &  & $<9.20$ & $>28.89$ & $>50.10$ & $>17.23$ & $<2.07$ & $>0.09$ & $>-1.23$\\
      \hline
      \multirow{2}{*}{AT2021aeou}  & 494 & \multirow{2}{*}{2.7 (fixed)} & $9.52\pm0.20$ & $29.45_{-0.11}^{+0.13}$ & $50.96\pm0.20$ & $17.09_{-0.24}^{+0.26}$ & $2.22_{-0.42}^{+0.41}$ & $0.12_{-0.05}^{+0.08}$ & $-1.12\pm 0.24$ \\
      & 786--808  & & $9.46_{-0.18}^{+0.13}$ & $29.10_{-0.08}^{+0.09}$ & $51.02_{-0.13}^{+0.18}$ & $16.98_{-0.15}^{+0.22}$ & $2.18_{-0.39}^{+0.27}$ & $0.06_{-0.02}^{+0.04}$ & $-0.47_{-0.23}^{+0.15}$\\
    \hline
    \end{tabular}
    \label{tab:22hvp_21aeou_radio_pars}
\end{table*}

The best-fit models are plotted in Figure~\ref{fig:radio_sed_22hvp_21aeou}, with model parameters ($\nu_{\rm p}$, $L_{\nu, \rm p}$, and $p$) presented in Table~\ref{tab:22hvp_21aeou_radio_pars}. Assuming that the outflow is launched around the time of optical first light, we can infer physical properties of the outflow and ambient environment.  
Following previous TDE radio studies \citep{Cendes2022_18hyz, Christy2024, Goodwin2025_eROSITA}, we adopt the equipartition ($\epsilon_{\rm e}=\epsilon_{B}=0.1$) derivations of \citet{BarniolDuran2013} and assume a spherical geometry of the outflow (cf., Eq.~4--13 in \citealt{Goodwin2022_19azh}). 
The equipartition energy ($E_{\rm eq}$) and radius ($R_{\rm eq}$) are the minimum values --- deviations from equipartition will render the parameters larger. 
The inferred parameters are shown in Table~\ref{tab:22hvp_21aeou_radio_pars}.
The inferred $E_{\rm eq}$, $\beta$, and $M_{\rm ej}$ are on the high end of those of X-ray (eROSITA) selected TDEs \citep{Goodwin2025_eROSITA}. We discuss the comparison with optically selected TDEs in \S\ref{subsec:extreme_disruption}.

\subsection{AT2020ysg and AT2020qhs}

For AT2020ysg and AT2020qhs, our observations are taken at late times ($t_{\rm rest}\sim2$--3\,yr). For simplicity, we assume that their outflow is in the Sedov–Taylor (ST) phase where energy is conserved\footnote{The TDE candidate eRASSt\,J234403-352640 entered the ST phase at $t_{\rm rest}\sim 500$\,d \citep{Goodwin2024}, and the TDE AT2019azh entered the ST phase at $t_{\rm rest}\sim 650$\,d \citep{Goodwin2022_19azh}. }. We note that our model does not apply if there are late-time energy injections from the TDE accretion flow. We prescribe the ambient density as $n(r) \propto r^{-k}$. In this case, the outflow velocity $\beta \propto t^{(k-3)/(5-k)}$, $L_{\nu,\rm p}$ and $\nu_{\rm p}$ can be solved as a function of $t_{\rm rest}$. Due to the paucity of data for these two sources, we fix $p = 2.7$. We assign flat priors: $0\leq k \leq 2.99$, $-3\leq {\rm log}\beta_{\rm 2.5\,yr} \leq -1$, where $\beta_{\rm 2.5\,yr}$ is the outflow velocity at $t_{\rm rest}=2.5$\,yr. For AT2020qhs, we further assume that the C-band light curve is in the optically thin regime, as is the case in other radio-detected TDEs with multi-band observations. 

\begin{table}[htbp!]
    \caption{Non-relativistic Sedov–Taylor Outflow modeling of AT2020ysg and AT2020qhs. \label{tab:20ysg_20qhs_radio_pars} }
    \centering
    \begin{tabular}{cccc}
    \hline\hline
       Name  &  $k$ & $E_{\rm eq}$ (erg)  \\
    \hline
       AT2020ysg  & $0.63_{-0.45}^{+0.74}$ & $10^{50.16\pm0.04}\left( \frac{\beta_{\rm 2.5\,yr}}{0.1}\right)^{1.407}$ \\
       AT2020qhs  & $0.68_{-0.48}^{+0.66}$  & $10^{50.33\pm0.05}\left( \frac{\beta_{\rm 2.5\,yr}}{0.1}\right)^{1.402}$\\
    \hline
    \end{tabular}
\end{table}

\begin{figure}[htbp!]
    \centering
    \includegraphics[width=0.55\linewidth]{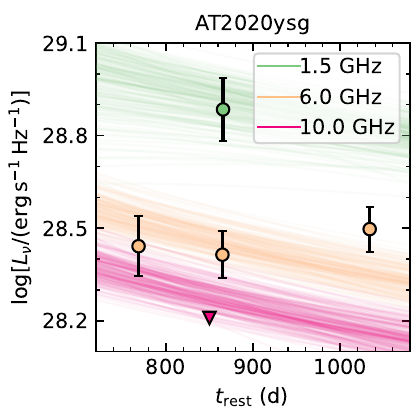}
    \includegraphics[width=0.435\linewidth]{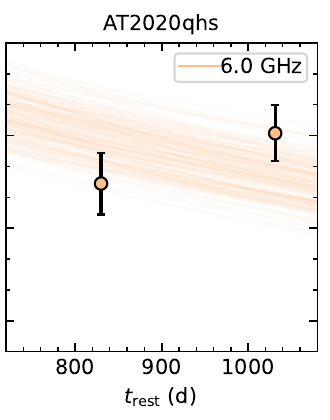}
    \caption{Radio light curves of AT2020ysg and AT2020qhs, overplotted with the best-fit Sedov–Taylor models. \label{fig:radio_lc_20ysg_20qhs}}
\end{figure}

The best-fit parameters are presented in Table~\ref{tab:20ysg_20qhs_radio_pars} and the corresponding models are overplotted on the data in Figure~\ref{fig:radio_lc_20ysg_20qhs}.
There exists a degeneracy between the inferred outflow velocity and energy.
The best-fit value of $k$ approaches the lower bound of the allowed range, reflecting the fact that the observed radio light curve remains flat rather than declining with time. 
This may indicate that the radio emitting region is experiencing a moderate amount of late-time energy injection from delayed outflows (not accounted for in our model) or pre-existing accretion activity (see \S \ref{subsec:agn_contribution}).

Assuming that the outflow entered into the ST phase at time $t_{\rm ST}$, we can compute the ejecta mass in the TDE outflow to be $M_{\rm ej} \sim 10^{-2.2}\,M_\odot \times ( \beta_{\rm 2.5\,yr}/0.03)^{-0.54} (t_{\rm ST}/{\rm 1\,yr} )^{1.09}$ for AT2020ysg 
and $M_{\rm ej} \sim 10^{-2.0}\,M_\odot \times ( \beta_{\rm 2.5\,yr}/0.03)^{-0.55} (t_{\rm ST}/{\rm 1\,yr} )^{1.07}$ for AT2020qhs.
We see that the amount of ejecta mass and total energy needed is within the budget expected in a TDE outflow. 

\section{Discussion}

\subsection{AGN contribution} \label{subsec:agn_contribution}


We discuss the possibility that the radio emission of AT2020ysg and AT2020qhs is powered by pre-existing active galactic nuclei (AGN)\footnote{To produce a radio luminosity of $\gtrsim 10^{38}\,{\rm erg\,s^{-1}}$ via star formation, the required star formation rate would be $\gtrsim 2.1\,M_\odot\,{\rm yr^{-1}}$, which is not consistent with host galaxy population synthesis analysis. Therefore, star formation is unlikely to be the source of the radio emission.}. 
Assuming a canonical AGN spectral shape of $f_\nu \propto \nu^{-0.7}$ \citep{Condon2002}, 
their 1.4\,GHz luminosity would be 
$\sim 10^{22.1}\,{\rm W\,Hz^{-1}}$, and their 150\,MHz luminosity would be 
$\sim 10^{22.8}\,{\rm W\,Hz^{-1}}$.
If the host galaxies of AT2020ysg and AT2020qhs indeed harbor AGN with such radio power, the absence of narrow emission lines in their optical spectra (see Appendix~\ref{subsec:20ysg_spec} and \citealt{Hammerstein2023}) would be consistent with expectations, as massive radio AGN host galaxies are typically quiescent systems \citep{Janssen2012, Jin2025_radio_AGN}.

Around $M_{\rm gal}\sim 10^{11}\,M_\odot$ (i.e., the typical host galaxy mass of our sample, see Figure~\ref{fig:Mgal_umr}), $\sim$3\% of galaxies harbor AGN with $L_{\nu, \rm 1.4\,GHz} \geq 10^{22.1}\,{\rm W\,Hz^{-1}}$ \citep{Best2005, Sabater2019}. This fraction shows no significant redshift evolution \citep{Kondapally2025}. 
Assuming that TDEs occur uniformly across all galaxy types, the probabilities of having zero, one, two and more than two out of eight TDEs occurring in radio-loud AGN is 78.4\%, 19.4\%, 2.1\%, and 0.1\%, respectively. Therefore, although we can not rule out the possibility that the detected radio emission of both AT2020ysg and AT2020qhs come from pre-existing AGN, this probability is low. 

Alternatively, the TDE rate may be enhanced in radio-bright galaxies.
Recent work by \citet{Kaur2025} investigates how the axisymmetric gravitational potential of a massive AGN disk can enhance TDE rates. In their model, the TDE rates are enhanced in AGN with massive gas disks, where $M_{\rm disk} \sim 0.1 M_{\rm BH}$. 
Similar to the standard scenario \citep{Stone2016}, most of the TDEs are channelled by highly eccentric orbits near the radius of influence of the central massive black hole: $r_{\rm h}\approx 10\,{\rm pc}(M_{\rm BH}/10^8\,M_\odot)[\sigma_\ast / (200\,{\rm km\,s^{-1}})]^{-2}$. 
However, jet-mode (radio-bright) AGN typically have much lower disk masses than radiative-mode AGN \citep{Yuan2014, Heckman2014}, potentially limiting this enhancement mechanism.

A different explanation for the preference of TDEs in jet-mode AGN may be related to black hole spin. 
Around the black hole mass of our sample ($M_{\rm BH}\sim 10^8\,M_\odot$; see Table~\ref{tab:sample}), a Sun-like star can be more easily disrupted if the BH spin is high \citep{Kesden2012, Huang2024, Mummery2024_spin}. Moreover, higher-spin black holes are more efficient at launching relativistic jets \citep{Tchekhovskoy2010}, providing a natural connection between enhanced spin, higher TDE rates, and stronger radio emission.

\subsection{No evidence of relativistic jets}\label{subsec:no_jets}

If any of the detected sources in our sample were associated with an off-axis relativistic jet launched at optical discovery with physical properties similar to those of on-axis jetted TDEs, we would expect one of the following observational signatures: (1) a declining light curve with luminosities comparable to those of on-axis jetted TDEs at similar epochs, or (2) a rising radio light curve (see Table~A1 of \citealt{Beniamini2023} for all the possible rise slopes. None of these signatures are observed. Therefore, there is no evidence for off-axis relativistic jets in our sample up to $t_{\rm rest}\sim 3\rm\, yrs$. 

Studies of known on-axis jetted TDEs showed that they are produced by black holes with $M_{\rm BH}\ll 10^8\rm\, M_\odot$, based on information about host galaxy, X-ray variability timescale, and jet turn-off time inferred from X-ray light curves (see discussion in \S3.4 of \citealt{Eftekhari2024}). However, it is possible that TDEs hosted by $\sim 10^8\,M_\odot$ black holes in our sample are preferentially highly spinning \citep{Kesden2012, Huang2024} which would favor formation of a powerful jet \citep{Tchekhovskoy2010}. Our non-detection of off-axis jets in our sample may be physically explained by the fact that the peak fallback rate of the stellar debris $\dot{M}_{\rm fb,peak}\propto M_{\rm BH}^{-1/2}$ drops below the Eddington limit $\dot{M}_{\rm Edd}\simeq 10 L_{\rm Edd}/c^2\propto M_{\rm BH}$ at high black hole mass, as the ratio between the two is roughly given by $\dot{M}_{\rm fb,peak}/\dot{M}_{\rm Edd}\simeq 1\, (M_*/M_\odot) (M_{\rm BH}/3\times10^7M_\odot)^{-3/2}$ for a main-sequence star of mass $M_*$ \citep{Law-Smith2020}. It has been argued that sub-Eddington, geometrically thin disks do not launch relativistic jets \citep{tchekhovskoy14_jetted_TDEs_MAD}. Our observations are consistent with this theoretical expectation. 

In Figure~\ref{fig:radio_lc}, we show example jetted TDE radio light curves for different viewing angle $\theta_{\rm obs}$, assuming all other model parameters follow the best-fit model inferred for Sw\,J1644 by \citet{Beniamini2023}. These other parameters include properties associated with the jet (isotropic equivalent kinetic energy $E_{\rm k,iso}$, initial Lorentz factor $\Gamma_0$, angular width $\theta_0$), microphysics ($p$, $\epsilon_e$, $\epsilon_B$), and the surrounding environment (the power-law index of the external density and its normalization). Note that the shape of the radio light curves could change a lot based on the assumed physical parameters. It is therefore possible that the radio emission from off-axis jets only stands out at later time $t_{\rm rest} > 3\rm\, yrs$. For instance, AT2018hyz, which has been proposed to harbor a powerful off-axis jet \citep{Matsumoto2023, Sfaradi2024}, showed dramatic late-time radio rebrightening with luminosities only becoming comparable to those of on-axis jets on timescales of several years. For this reason, we encourage follow-up observations of the radio detected sources in our sample at later epochs.

\subsection{Association with more energetic prompt outflows} \label{subsec:extreme_disruption}

We would like to ask: statistically speaking, are optically overluminous TDEs also more luminous in the radio band? If so, this will indicate that optically overluminous TDEs might be associated with more energetic prompt\footnote{``Prompt'' refers to a launch time that is near the optical first light epoch.} outflows. The inferred values of $E_{\rm eq}\sim 10^{50}$--$10^{51}$\,erg and $\beta\sim 0.1$ for AT2022hvp and AT2021aeou (see Table~\ref{tab:22hvp_21aeou_radio_pars}) lie at the high end of the distribution for TDEs with radio emission attributed to prompt outflows, where $10^{47}\lesssim (E_{\rm eq}/{\rm erg})\lesssim 10^{51}$ and $0.01\lesssim \beta\lesssim 0.1$ (cf., Fig.~5 of \citealt{Goodwin2025_eROSITA}). 

To address this question, we compare our radio detections and upper limits obtained at $1 < (t_{\rm rest}/{\rm yr}) < 2$ for AT2022hvp, AT2021aeou, AT2021gje, AT2020acka, and AT2021yzv with previously known fainter ($M_{g,\rm peak}>-20.5$\,mag) events at similar phases. We did not include AT2020qhs, AT2020ysg, and AT2019cmw in this analysis, as (1) they have observations conducted at $>2$\,yr, which is more sensitive to delayed outflow launch \citep{Cendes2024}, and (2) the origin of the radio emission in AT2020qhs and AT2020ysg is ambiguous (\S\ref{subsec:agn_contribution}). 

\begin{figure}[htbp!]
    \centering
    \includegraphics[width=\columnwidth]{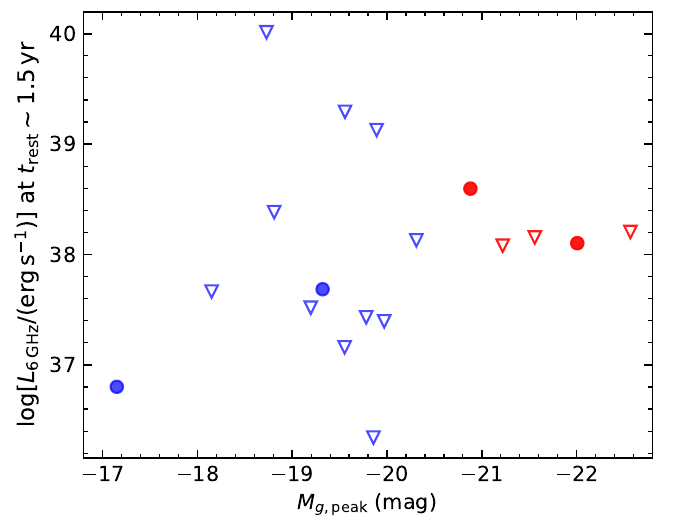}
    \caption{Radio luminosities at $t_{\rm rest} \sim1.5$\,yr for optically overluminous TDEs ($M_{g, \rm peak}<-20.8$; shown in red) and a control sample of fainter TDEs ($M_{g, \rm peak}>-20.5$; shown in blue). Considering both detections (solid circles) and upper limits (hollow downward triangles), our Bayesian analysis shows that our sample is brighter than the control sample at $2.1\sigma$ significance.
    \label{fig:compare}}
\end{figure}

We construct a comparison sample of 13 TDEs with $M_{g, \rm peak}>-20.5$ from \citet{Cendes2024} (see Appendix~\ref{sec:comparison} for details). Figure~\ref{fig:compare} shows the $\sim6$\,GHz radio luminosities and upper limits of our sample and the comparison sample. 

To assess whether the radio luminosity distributions of the two samples differ statistically, we performed a Bayesian analysis that accounts for censored data (i.e., upper limits). 
We modeled the logarithmic radio luminosities in each sample as being drawn from independent normal distributions, characterized by population means $\mu_1$ and $\mu_2$ and a shared standard deviation $\sigma$. 
Here, $\mu_1$ corresponds to the comparison sample (dataset 1), and $\mu_2$ to our sample of optically overluminous TDEs (dataset 2). 
We assumed normal priors for $\mu_1$ and $\mu_2$, each centered at 38 with a standard deviation of 1 dex, and a half-normal prior on the shared standard deviation $\sigma$ with a scale of 1 dex. 
Upper limits were treated as censored observations by including a cumulative log-probability term 
\begin{align}
    {\rm log} P (x < x_{\rm lim}) = {\rm log} \Phi \left( \frac{x_{\rm lim} - \mu}{\sigma} \right)
\end{align}
in the model likelihood, where $\Phi$ is the normal cumulative distribution function. The posterior distributions were inferred using MCMC sampling as implemented in \texttt{PyMC} \citep{pymc}. 

The posterior distribution of the mean difference, $\delta_\mu = \mu_1 - \mu_2$, is centered at $-1.2$ dex, with a 94\%\footnote{This is the default HDI value in the Bayesian visualization package \texttt{ArviZ} \citep{arviz_2019}. 94\% is chosen as a balance between being informative and not overly wide.} highest density interval (HDI) spanning [$-2.1$, $-0.18$]. The posterior probability that $\delta_\mu > 0$ is 1.6\%, which corresponds to a one-sided 2.1$\sigma$ level of confidence in favor of $\delta_\mu < 0$. 

This result tentatively supports a connection between high optical peak luminosity and stronger radio emission on the timescale of $\sim1.5$\,yr. It further suggests that overluminous TDEs may be associated with more energetic prompt outflows. 

A likely explanation involves the high black hole mass ($M_{\rm BH}\sim 10^8\,M_\odot$; see Table~\ref{tab:sample}) found in these overluminous events. Since the tidal radius $r_{\rm T} \sim R_\ast (M_{\rm BH} / M_\ast)^{1/3}$, whereas the gravitational radius $r_{\rm g} = GM_{\rm BH} / c^2$, we have $r_{\rm T} / r_{\rm g} \sim 10 (M_{\rm BH} / 10^7\,M_\odot)^{-2/3}$ for a Sun-like star. Therefore, general relativistic effects become increasingly important for disruption by higher-mass black holes. 
In particular, strong relativistic apsidal precession can cause enhanced mixing and collisions between the bound and unbound debris streams \citep{Laguna1993, Cheng2014, Gafton2019, Ryu2023_extremeTDE}. 
These interactions promote the circularization of bound material via stream-stream collision, while also energizing the unbound debris.
The source of the outflows responsible for radio emission may be caused by either the stream-stream collision  \citep{Lu2020}, or the unbound debris \citep{Krolik2016}. In the latter case, \citet{Ryu2023_extremeTDE} showed that in extreme TDEs (defined by pericenter distances $r_{\rm p} < 6 r_{\rm g}$), up to $\sim$1\% of the unbound debris's mass can reach velocities exceeding $2\times 10^4\,{\rm km\,s^{-1}}$, significantly faster than in typical TDEs hosted by lower-mass black holes.

At the same time, theoretical studies have shown that, all else being equal (e.g., black hole spin and orbital inclination), the Hills mass is larger for higher-mass stars \citep{Huang2024, Mummery2024_spin}. This raises the possibility that the stars disrupted by $M_{\rm BH} \sim 10^8\,M_\odot$ black holes are themselves more massive. In such cases, the disrupted star can provide a larger energy reservoir, potentially powering a more energetic outflow and hence stronger radio emission. Future demographics studies are needed to test this hypothesis.

\section{Conclusion}

In this paper we have presented VLA radio observations of eight optically overluminous TDEs ($M_{g, \rm peak}<-20.8$) at $1<(t_{\rm rest}/{\rm yr}) < 3$. The host galaxies of our sample are massive ($10^{10.6}\lesssim M_{\rm gal}\lesssim 10^{11.2}$), and the black hole masses ($10^{7.7}\lesssim M_{\rm BH}\lesssim 10^{8.3}$) are on the high end of the TDE population \citep{Yao2023}.

We summarize the main conclusions below:
\begin{itemize}
    \item Radio emission is not detected in four out of the eight TDEs in our sample. 
    \item Among the four detected TDEs, their radio luminosities are in the range of $10^{38}$--$10^{39}\,{\rm erg\,s^{-1}}$ and around two orders of magnitude fainter than those of on-axis jetted TDEs at the same phase.
    \item We find no evidence for off-axis jets in our sample, disfavoring a connection between optically overluminous TDEs and off-axis relativistic jets.
    \item Among the detected events, AT2022hvp and AT2021aeou show clear spectral evolution in their radio SEDs, while AT2020ysg and AT2020qhs exhibit no statistically significant variability.
    \item The radio evolution of AT2022hvp and AT2021aeou at $1<t_{\rm rest}<2$\,yr is consistent with synchrotron emission from non-relativistic outflows launched at the epoch of first optical light. The inferred equipartition energy is $E_{\rm eq} \sim 10^{50}$--$10^{51}\,{\rm erg\,s^{-1}}$, with outflow velocities of $\beta \sim 0.1c$ --- these values are on the high end of TDE prompt outflow properties that have been inferred using similar approaches. 
    \item The radio emission of AT2021aeou and AT2020qhs at $2<t_{\rm rest}<3$\,yr may originate either from pre-existing AGN activity or from non-relativistic outflows launched by the TDEs themselves. If the former is true, our results suggest an enhanced TDE rate in radio-loud AGN. 
    \item Five TDEs in our sample have radio observations at $t_{\rm rest} \sim 1.5$\,yr. When compared to the 6\,GHz luminosities (including both detections and upper limits) of opticaller fainter TDEs with $M_{g, \rm peak}> -20.5$\,mag, these overluminous TDEs exhibit systematically brighter radio emission, with a statistical significance of $2.1\sigma$.
    This result supports the idea that overluminous TDEs generate more energetic non-relativistic outflows, potentially driven by strong general relativistic effects during disruptions by higher-mass black holes or the disrupted star being more massive. 
\end{itemize}

Continued sensitive radio observations of this sample are needed to rule out the presence of off-axis jets with radio light curves that may rise on timescales $\gtrsim 3$\,yr, and to determine the true origin of the radio emission in AT2020ysg and AT2020qhs. 
In addition, targeted radio monitoring of a larger sample of optically bright TDEs, with well-designed cadences, will be essential to confirm or refute the statistical trend between optical and radio luminosities.

\begin{acknowledgments}
YY would like to thank Karamveer Kaur for helpful discussions regarding TDE rates.

Based on observations obtained with the Samuel Oschin Telescope 48-inch and the 60-inch Telescope at the Palomar Observatory as part of the Zwicky Transient Facility project. ZTF is supported by the National Science Foundation under Grants No. AST-1440341, AST-2034437, and currently Award \#2407588. ZTF receives additional funding from the ZTF partnership. Current members include Caltech, USA; Caltech/IPAC, USA; University of Maryland, USA; University of California, Berkeley, USA; University of Wisconsin at Milwaukee, USA; Cornell University, USA; Drexel University, USA; University of North Carolina at Chapel Hill, USA; Institute of Science and Technology, Austria; National Central University, Taiwan, and OKC, University of Stockholm, Sweden. Operations are conducted by Caltech's Optical Observatory (COO), Caltech/IPAC, and the University of Washington at Seattle, USA.

The Gordon and Betty Moore Foundation, through both the Data-Driven Investigator Program and a dedicated grant, provided critical funding for SkyPortal.
The ZTF forced-photometry service was funded under the Heising-Simons Foundation grant No. 12540303 (PI: Graham).

The National Radio Astronomy Observatory is a facility of the National Science Foundation operated under cooperative agreement by Associated Universities, Inc.
\end{acknowledgments}


\facilities{VLA, Keck:I (LRIS), Keck:II (ESI), Hale, PO:1.2m}

\software{\texttt{ArviZ} \citep{arviz_2019},
\texttt{astropy} \citep{Astropy2022}, 
\texttt{CASA} \citep{CASATeam2022},
\texttt{emcee} \citep{Foreman-Mackey2013},
\texttt{LPipe} \citep{Perley2019lpipe}, 
\texttt{matplotlib} \citep{Hunter2007},
\texttt{PyMC} \citep{pymc}
}


\appendix

\section{Velocity Dispersion and Black Hole Mass} \label{sec:velocity_dispersion}

\begin{deluxetable*}{lcccccccc}[htbp!]
\tabletypesize{\small}
\tablecaption{Log of medium-resolution optical spectroscopy with Keck-II ESI. \label{tab:spec_medres}}
\tablehead{
    \colhead{IAU Name}
    & \colhead{Start Date}  
	& \colhead{$t_{\rm rest}$ (days)}
	& \colhead{Fitted $\lambda_{\rm rest}$ (\AA)\tablenotemark{a}}
	& \colhead{Slit Width ($^{\prime\prime}$)}
	& \colhead{Exp. (s)} 
        & \colhead{$r_{\rm extract}$ (pixel)\tablenotemark{b}} 
        & \colhead{S/N}
	}
\startdata
AT2021gje & 2022-03-07.6 & 298 &  3900--5300\tablenotemark{a} & 0.5 & 1800 & 3.2 & 6.6\\
AT2022hvp & 2022-11-27.6 & 210 & 5030--5600\tablenotemark{a}  & 0.75  & 1800 & 5.3 & 10.4\\ 
\enddata 
\tablenotetext{a}{Wavelength range used for spectral fitting.}
\tablenotetext{b}{The radius used for extracting the spectrum. $r_{\rm extract}$ can be converted to angular scale using a conversion factor of 0.154$^{\prime\prime}$ per pixel.}
\end{deluxetable*}

\begin{figure*}[htbp!]
    \centering
    \includegraphics[width=\textwidth]{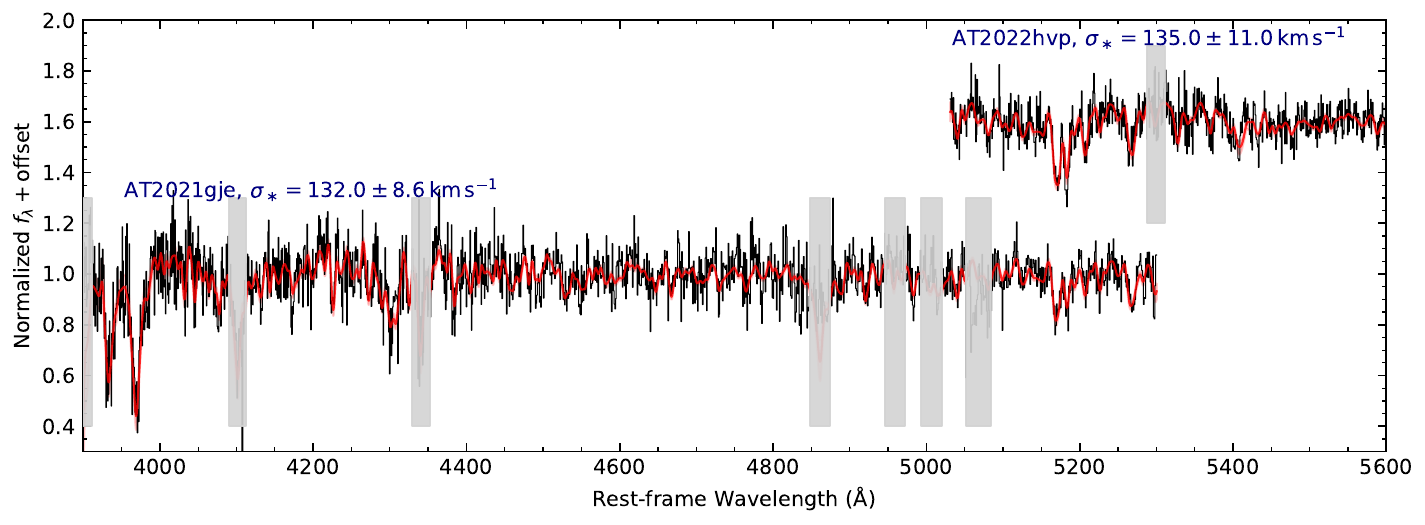}
    \caption{ESI spectrum of the host galaxies of AT2021gje and AT2022hvp (black) and the best-fit models (red).\label{fig:esi} }
\end{figure*}

In Table~\ref{tab:sample}, the velocity dispersion $\sigma_\ast$ measurements for host galaxies of AT2020acka, AT2020ysg, and AT2021yzv are taken from \citet{Yao2023}, and for AT2020qhs is taken from \citet{Hammerstein2023_KCWI}. 
We obtained additional medium-resolution spectroscopy for the host galaxies of AT2021gje and AT2022hvp using the Echellette Spectrograph and Imager (ESI; \citealt{Sheinis2002}) on the Keck II telescope. 
We follow the same procedures as outlined in \citet{Yao2023} to reduce the data and measure $\sigma_\ast$ using the penalized pixel-fitting (\texttt{pPXF}) software \citep{Cappellari2004, Cappellari2017}. For AT2022hvp, we adopt the same rest-frame wavelength range of 5030--5600\,\AA\ for spectral fitting. For AT2021gje, due to the lower S/N, we adopt a wider spectral range to include the \ion{Ca}{II} H and K lines into the fitting. The data and best-fit models are shown in Figure~\ref{fig:esi}. 

We estimate the black hole mass $M_{\rm BH}$ using host galaxy scaling relations, including the $M_{\rm BH}$--$\sigma_\ast$ relation \citep{Kormendy2013} for six host galaxies with $\sigma_{\ast}$ measurements, and the $M_{\rm BH}$--$M_{\rm gal}$ relation (derived by \citealt{Greene2020} using all types of galaxies) for the other two TDE hosts.

\section{Low-resolution Optical Spectroscopy of Selected TDEs} \label{sec:opt_spec}

\begin{deluxetable*}{lccccccc}[htbp!]
\tabletypesize{\small}
\tablecaption{Log of low-resolution optical spectroscopy. \label{tab:spec_lowres}}
\tablehead{
    \colhead{IAU Name}
    & \colhead{Start Date}  
	& \colhead{$t_{\rm rest}$ (days)}
	& \colhead{Telescope}
	& \colhead{Instrument}
	& \colhead{Wavelength Range (\AA)}
	& \colhead{Slit Width ($^{\prime\prime}$)}
	& \colhead{Exp. (s)} 
	}
\startdata
AT2022hvp & 2022-05-04.1 & 24 & P200 & DBSP & 3410–5550, 5750–9995 & 1.5 & 450 \\
AT2022hvp & 2022-05-23.2 & 41 & P200 & DBSP & 3410–5550, 5750–9995 & 1.5 & 300 \\
AT2022hvp & 2022-05-26.3 & 44 & Keck I & LRIS & 3260--10250 & 1.0 & 600 \\
AT2022hvp & 2022-10-31.6 & 186 & Keck I & LRIS & 3200--10250 & 1.0 & 660/600\tablenotemark{a} \\
\hline
AT2020ysg & 2020-12-12.6 & 141 & Keck I & LRIS & 3200--10250 & 1.0 & 300 \\
AT2020ysg & 2023-04-21.5 & 815 & Keck I & LRIS & 3200--10250 & 1.0 & 1760 \\
\hline
AT2021aeou & 2022-07-02.2 & 250 & Keck I & LRIS & 3200--10250 & 1.0 & 1200\\
AT2021aeou & 2022-08-02.3 & 264 & Keck I & LRIS & 3200--10250 & 1.0 & 900 \\
AT2021aeou & 2023-04-21.5 & 460 & Keck I & LRIS & 3200--10250 & 1.0 & 1760 \\
\enddata 
\tablenotetext{a}{Exposure times on blue/red sides of the spectrograph.}
\end{deluxetable*}

Here we provide additional follow-up spectra obtained by us for this sample (see Table~\ref{tab:spec_lowres} for a log), using the DBSP on the Palomar 200-inch Hale telescope (P200), and the Low Resolution Imaging Spectrograph (LRIS; \citealt{Oke1995}) on the Keck-I telescope. 
These observations were coordinated using the \textit{fritz.science} instance of \texttt{SkyPortal} \citep{vanderWalt2019, Coughlin2023}.
Instrumental setup and data reduction of DBSP and LRIS spectra are the same as outlined in Appendix~B of \citet{Yao2022_21ehb}. 

Based on these, we change the spectral subtype of AT2020ysg from TDE-featureless to TDE-He (\S\ref{subsec:20ysg_spec}), classify AT2022hvp as TDE-He (\S\ref{subsec:22hvp_spec}), and AT2021aeou as TDE-featureless (\S\ref{subsec:21aeou_spec}). 

\subsection{AT2020ysg} \label{subsec:20ysg_spec}

\begin{figure}[htbp!]
    \centering
    \includegraphics[width = \columnwidth]{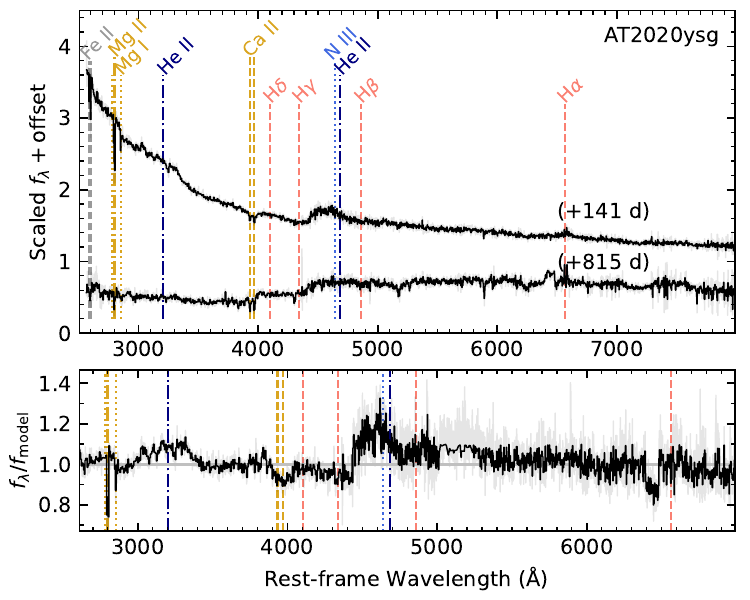}
    \caption{\textit{Upper}: LRIS spectra of AT2020ysg.
    \textit{Bottom}: The +141\,d spectrum divided by a blackbody+host model, which shows signatures of broad emission lines around \ion{He}{II}. \label{fig:20ysg_opt_spec}}
\end{figure}

\citet{Hammerstein2023} presented two LDT/DeVeny spectra of AT2020ysg obtained on 2020 December 6 ($t_{\rm rest}=136$\,d) and 2021 January 11 ($t_{\rm rest}=165$\,d). Since these spectra show blue continua and no discernible broad emission features, AT2020ysg was classified as TDE-featureless in \citet{Hammerstein2023}.

Here, in the upper panel of Figure~\ref{fig:20ysg_opt_spec}, we show two LRIS spectra of AT2020ysg obtained at $t_{\rm rest}=141$\,d and 815\,d. The 815\,d spectrum was obtained at a sufficiently late time such that it is dominated by the host galaxy light. This allows us to reveal weak TDE spectral features in the 141\,d data by comparing it with the blackbody+host model, where the host contribution is given by the 815\,d spectrum. The data divided by model is shown in the bottom panel of Figure~\ref{fig:20ysg_opt_spec}, where broad \ion{He}{II} emission lines at $\lambda 3203$ and $\lambda 4686$ are evident. We note that the \ion{He}{II} $\lambda 3203$ line is also seen in the most recent analysis of the 165\,d LDT spectrum by \citet{Hammerstein2025}. The \ion{He}{II} $\lambda 4686$ line profile looks blueshifted and asymmetric, similar to the early-time behavior of ASASSN-15oi \citep{Holoien2016_15oi}. As such we change the spectral subtype of this object from TDE-featureless to TDE-He.

\subsection{AT2022hvp} \label{subsec:22hvp_spec}

\begin{figure}[htbp!]
    \centering
    \includegraphics[width = \columnwidth]{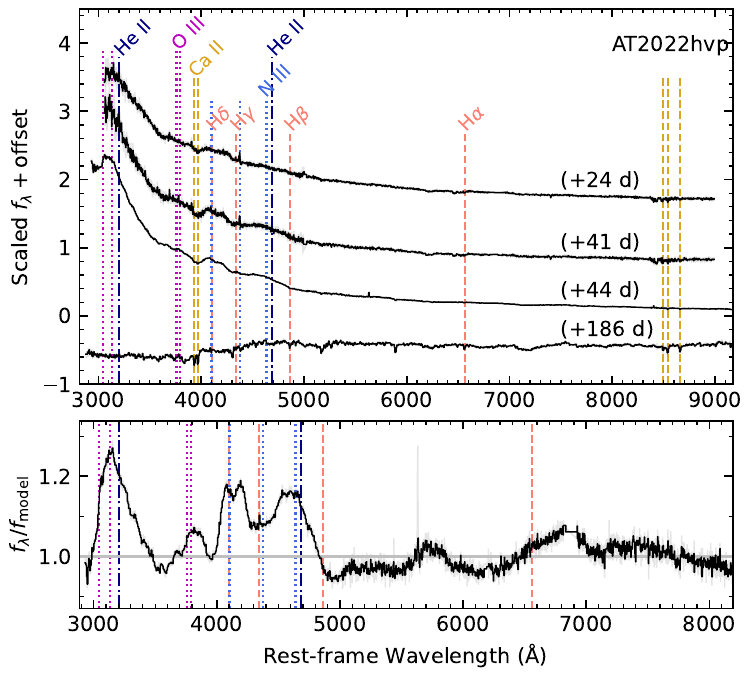}
    \caption{\textit{Upper}: DBSP and LRIS spectra of AT2022hvp.
    \textit{Bottom}: The +44\,d spectrum divided by a blackbody+host model, which shows signatures of broad emission lines around \ion{O}{III}, \ion{N}{III}, and \ion{He}{II}. 
    \label{fig:22hvp_opt_spec}}
\end{figure}

AT2022hvp was classified as a TDE-He at $z=0.12$ using a spectrum obtained on 2022-05-06.9 ($t_{\rm rest}=27$\,d) by LT/SPART \citep{Fulton2022_22hvp}, which displays a blue continuum and broad emission features around \ion{He}{II} and \ion{N}{III}. The redshift of 0.12 comes from template matching. 
In the upper panel of Figure~\ref{fig:22hvp_opt_spec}, we show the DBSP and LRIS spectra of AT2022hvp obtained by us. We assign a more accurate redshift of $z=0.112$ for this TDE, as absorption lines (e.g., \ion{Ca}{II}, \ion{H}{I}, \ion{Na}{I}) from the host galaxy are identified at this redshift. 

Since the +44\,d LRIS spectrum was obtained around maximum light where transient dominates, 
we attempt to reveal weaker line features by comparing it with a blackbody+host model, where the host is approximated by the +186\,d LRIS spectrum. 
In the bottom panel of Figure~\ref{fig:22hvp_opt_spec}, we present the data divided by model, which shows four broad emission feature. 
The feature around \ion{He}{II} $\lambda4686$/\ion{N}{III} $\lambda\lambda4634,4641$ is commonly seen in optically selected TDEs. 
The feature at H$\delta$/\ion{N}{III} $\lambda4104$ most likely come from \ion{N}{III}, as there is no signatures of lower-order Balmer series. 
The feature at $\sim 3800$\AA\ can be attributed to the \ion{O}{III} $\lambda\lambda$3760, 3791 Bowen fluorescence lines \citep{Selvelli2007}.
The feature at $\sim3100$\AA\ can be attributed to a combination of \ion{He}{II} $\lambda 3203$ and the \ion{O}{III} $\lambda\lambda$3047, 3133 Bowen fluorescence lines. 

We note that the intrinsic line ratio of \ion{He}{II} $\lambda3203/\lambda4686$ is expected to be 0.45 under Case B recombination conditions \citep{Storey1995, Gezari2012}. The apparently enhanced strength of the \ion{He}{II} $\lambda3203$ line may reflect complexities in the ionization structure of the reprocessing envelope \citep{Roth2016}, a detailed investigation of which is beyond the scope of this work.

\subsection{AT2021aeou} \label{subsec:21aeou_spec}

\begin{figure}[htbp!]
    \centering
    \includegraphics[width = \columnwidth]{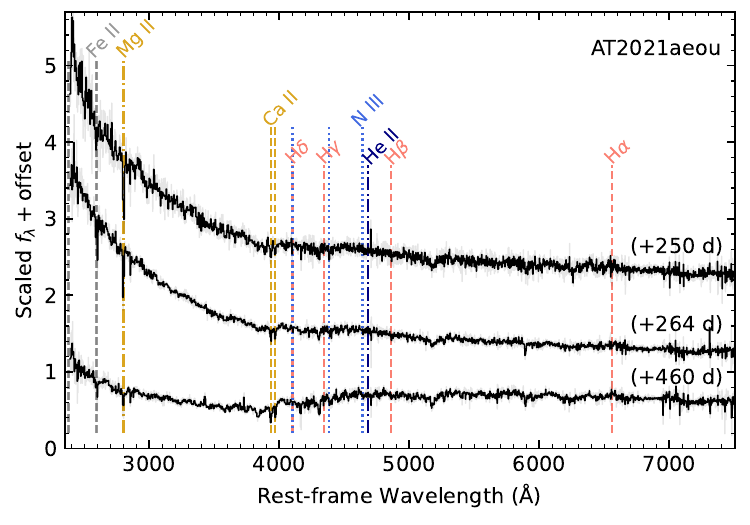}
    \caption{LRIS spectra of AT2021aeou. \label{fig:21aeou_opt_spec}}
\end{figure}

Figure~\ref{fig:21aeou_opt_spec} shows the LRIS spectra of AT2021aeou. We note that the optical light curve of this object exhibits a double-peaked profile (see the cyan circles in the bottom panel of Figure~\ref{fig:lumTDE_lc}). All LRIS spectra presented here were obtained during the decline of the second peak. The transient exhibits featurless spectra. 

\section{Additional Tables} \label{sec:tables}
The VLA observing log is shown in Table~\ref{tab:vla}.

\begin{deluxetable*}{l|c|c|c|c|c|c}
	\tablecaption{Targeted VLA observations of overluminous TDEs.\label{tab:vla}}
	\tablehead{
        \colhead{IAU name}
		& \colhead{Date}   
		& \colhead{$t_{\rm rest}$ }
		& \colhead{Receiver} 
		& \colhead{$\nu_0$} 
		& \colhead{$f_\nu$}
        & \colhead{Array Configuration}\\
        \colhead{}   
		& \colhead{}   
		& \colhead{(day)}
		& \colhead{} 
		& \colhead{(GHz)}
		& \colhead{($\mu$Jy)}
        & \colhead{} 
	}
	\startdata
 \multirow{5}{*}{AT2020ysg} & 2023-02-22.2 & 769  & C & 5.999 & $14.7\pm3.3$ & B\\
 \cline{2-7}
                            & 2023-06-05.9 & \multirow{3}{*}{865} & X & 9.999     & $<8.6$ & BnA \\
 \cline{4-6}
                            & 2023-06-24.9 &   & C & 5.999 & $13.8\pm 2.4$ & BnA $\rightarrow$ A \\
 \cline{4-6}
                            & 2023-06-25.8 &   & L &  1.520  & $40.8\pm9.7$ & BnA $\rightarrow$ A \\
 \cline{2-7}
                            & 2024-01-26.3 &  1034 & C &  5.999  & $16.7\pm2.8$ & C \\
 \hline 
 AT2021yzv & 2023-03-12.0 & 451 & C & 5.999 & $<9.2$ & B\\
 \hline 
 \multirow{8}{*}{AT2021aeou} & 2023-03-18.3 & 435 & C & 5.999  & $37.0\pm7.6$ & B \\
 \cline{2-7}
  & 2023-06-05.9 & \multirow{3}{*}{494}& L & 1.520 & $<62.1$ & \multirow{3}{*}{BnA}\\
  & 2023-06-05.9 & & C & 5.999 & $41.0 \pm 7.7$ &\\
  & 2023-06-06.0 & & X & 9.999 & $27.4\pm 6.2$ &\\
  \cline{2-7}
  & 2024-07-01.1 & 786 & L & 1.520 & $<29.1$ & B\\
  \cline{2-7}
  & 2024-07-30.1 & \multirow{3}{*}{808} & S & 2.999 & $18.9\pm4.5$ & \multirow{3}{*}{B}\\
  & 2024-07-30.0 &  & C & 5.999 & $13.9\pm4.1$ & \\
  & 2024-07-30.0 &  & X & 9.999 & $13.8\pm4.6$ &\\
 \hline 
 AT2020acka & 2023-03-21.4 & 672 & C & 5.999  & $<7.0$ & B\\
 \hline 
 AT2019cmw & 2023-03-30.4 & 971  & C & 5.999  & $<4.5$ & B\\
 \hline 
 AT2021gje & 2023-04-02.3 & 586  & C & 5.999  & $<4.6$ & B\\
 \hline
 \multirow{27}{*}{AT2022hvp} & 2022-06-11.0 & 58 & Ku & 15.000 & $<16.2$ & A\\
  \cline{2-7}
    & \multirow{2}{*}{2022-11-23.7} &\multirow{2}{*}{207}   & \multirow{2}{*}{X} & 11.000 & $121.8\pm 12.7$ &\multirow{2}{*}{C} \\
                             &  &  &  & 9.000 & $154.8 \pm 13.0 $ & \\
\cline{2-7}
    & \multirow{3}{*}{2023-03-12.3} & \multirow{3}{*}{305}  & Ku & 15.000 & $58.3\pm 9.7$ & \multirow{3}{*}{B}\\
  \cline{4-6}
    &  &  &  \multirow{2}{*}{X} & 11.000 & $107.6\pm20.3$ & \\
    &  &  &  & 9.000 & $131.5\pm 18.7$ & \\
 \cline{2-7}
  & \multirow{6}{*}{2023-04-08.1} & \multirow{6}{*}{329}  & \multirow{6}{*}{C} & 4.103 & $209.6\pm15.0$ & \multirow{6}{*}{B}\\
 & & & & 4.359 & $205.5\pm12.0$ &\\
 & & & & 4.743 & $200.8\pm8.5$ &\\
 & & & & 5.511 & $171.8\pm8.3$ &\\
 & & & & 6.487 & $155.9\pm6.7$ &\\
 & & & & 7.511 & $129.2\pm6.3$ &\\
   \cline{2-7}
  & \multirow{6}{*}{2023-04-22.0} & \multirow{6}{*}{341}  &  \multirow{2}{*}{L} & 1.264 & $<211.8$ & \multirow{6}{*}{B}\\
                                                                      & & & & 1.776 & $162.6 \pm 33.6$ & \\
                                                        \cline{4-6}
  & & & \multirow{2}{*}{S} & 2.499 & $246.0\pm 24.1$ & \\
  & & & & 3.499 & $173.4 \pm 20.5$ &  \\
  \cline{4-6}
  & & & \multirow{2}{*}{C} &4.999 & $148.8 \pm 20.4$ & \\
  & & & & 6.999 & $143.4 \pm 15.5$  &  \\
  \cline{2-7}
  & \multirow{5}{*}{2024-01-25.4} &   \multirow{5}{*}{592} & L & 1.519 & $<498.0$ & \multirow{5}{*}{C} \\
  \cline{4-6}
   & & & S & 2.999 & $135.6 \pm 27.3$ & \\ 
   \cline{4-6}
   & & & \multirow{2}{*}{C} & 4.871 & $101.2 \pm 16.1$ & \\
   & & &  & 6.959 & $68.3 \pm 15.2$ & \\
   \cline{4-6}
   & & & X & 9.999 & $55.3\pm7.1$ & \\
   \cline{2-7}
   &  \multirow{4}{*}{2024-09-08.6} & \multirow{4}{*}{796} & L & 1.52 & $215\pm 53$ & \multirow{4}{*}{B}\\
   & & & S & 3.00 & $107\pm21$ & \\
   & & & C & 6.00 & $64\pm11$  & \\
   & & & X & 10.00 & $40\pm 10$ & \\
   \hline 
 \multirow{2}{*}{AT2020qhs} & 2023-04-28.9  & 830  & C & 5.999  & $14.7\pm 3.4$ & B\\
 \cline{2-7}
        & 2024-01-25.1 & 1034 & C & 6.000 & $21.4\pm4.5$ & C\\
\enddata
\tablecomments{$\nu_0$ is observed central frequency. $f_\nu$ is the observed flux density values. }
\end{deluxetable*}

\section{The Comparison Sample} \label{sec:comparison}

Our goal is to compare the $\sim 6$\,GHz radio luminosities of over-luminous TDEs with those of TDEs with $M_{g, \rm peak}>-20.5$\,mag at similar post-disruption phases. For the former, we use five TDEs in our sample, each with 6\,GHz observations at $1.2<t_{\rm rest}<1.9$\,yr. For the latter, we adopt the comparison sample from \citet{Cendes2024}, which presents late-time radio follow-up of a systematically selected TDE sample. Unless otherwise noted, the data are taken from \citet{Cendes2024}.

Among the 24 TDEs in the \citet{Cendes2024} sample, we exclude:
\begin{itemize}
    \item Six events (OGLE17aaj, AT2018bsi, AT2020mot, AT2020nov, AT2020pj, and AT2020wey) whose radio emission has ambiguous origins or is likely dominated by AGN activity.\footnote{Five of them have observations at $\sim 1.5$\,yr. If we include these events and treat them as upper limits, the significance of our result in \S\ref{subsec:extreme_disruption} increases from 2.1$\sigma$ to $2.5\sigma$.}
    \item Three events (ASASSN-14ae, iPTF16axa and AT2018lna) that lack radio observations before 2\,yr.
    \item One event (PS16dtm) with a luminous peak of $M_{g, \rm peak}\sim -21.8$ \citep{Blanchard2017, Petrushevska2023}.
    \item One event (DES14C1kia) that do not have a publicly available peak optical magnitude. 
\end{itemize}

This leaves a comparison sample of 13 TDEs. Ten of them have $\sim6$\,GHz constraints around $t_{\rm rest} \sim 1.5$\,yr, including 2 with detections and 8 with upper limits. 

Two events with $\sim1.5$\,yr detections:
\begin{enumerate}
    \item iPTF16fnl: Detected at 15.5\,GHz with a flux of 151\,$\mu$Jy at $t_{\rm rest} = 1.14$\,yr \citep{Horesh2021}, corresponding to a 6\,GHz flux of 338\,$\mu$Jy assuming an optically thin spectrum with $p = 2.7$. A 6\,GHz detection at 45\,$\mu$Jy exists at $t_{\rm rest} = 3.69$\,yr. Interpolating in log-log space gives a 6\,GHz flux of 211\,$\mu$Jy at $t_{\rm rest} = 1.5$\,yr.
    \item AT2019dsg: Detected at 6\,GHz with a flux of 130\,$\mu$Jy at $t_{\rm rest} = 1.49$\,yr \citep{Cendes2021_19dsg}.
\end{enumerate}

Eight events with interpolatable $\sim1.5$\,yr upper limits:
\begin{enumerate}
    \item AT2018zr: 10\,GHz non-detection at $<37.5$\,$\mu$Jy ($t_{\rm rest} = 0.19$\,yr; \citealt{vanVelzen2019}) and 6\,GHz non-detection at $<14$\,$\mu$Jy ($t_{\rm rest} = 2.36$\,yr). We adopt a 6\,GHz upper limit of $<30$\,$\mu$Jy at $t_{\rm rest} \sim 1.5$\,yr.
    \item AT2018dyb: 19\,GHz upper limit of $<43$\,$\mu$Jy at $t_{\rm rest} = 0.09$\,yr \citep{Holoien2020}; 1.36\,GHz detection at 158\,$\mu$Jy at $t_{\rm rest} = 2.78$\,yr implies 6\,GHz flux of 45\,$\mu$Jy (assuming $p = 2.7$). As the light curve rises at later times, we adopt a conservative 6\,GHz upper limit of $<50$\,$\mu$Jy at 1.5\,yr.
    \item AT2018hyz: VLASS 3\,GHz upper limit of $<0.45$\,mJy at $t_{\rm rest} = 1.84$\,yr \citep{Cendes2022_18hyz}.
    \item AT2019eve: VLASS 3\,GHz upper limit of $<0.497$\,mJy at $t_{\rm rest} = 1.33$\,yr.
    \item AT2020neh: 15\,GHz upper limit of $<16$\,$\mu$Jy at $t_{\rm rest} = 0.51$\,yr \citep{Angus2022}; 6\,GHz detection at 26\,$\mu$Jy at $t_{\rm rest} = 2.25$\,yr. Given the rising light curve at later times, we adopt a 6\,GHz upper limit of $<26$\,$\mu$Jy at 1.5\,yr.
    \item iPTF15af: 6\,GHz non-detections at $t_{\rm rest} = 0.04$ and 4.6\,yr. We adopt an upper limit of $<50$\,$\mu$Jy at 1.5\,yr.
    \item AT2017eqx: 6\,GHz non-detections at $t_{\rm rest} = 0.11$\,yr \citep{Nicholl2019} and 2.69\,yr. We adopt an upper limit of $<18$\,$\mu$Jy at 1.5\,yr.
    \item AT2018fyk: 19\,GHz non-detection of $<53$\,$\mu$Jy at $t_{\rm rest} = 0.32$\,yr \citep{Wevers2019_18fyk}, and 1.36\,GHz upper limit of $<60$\,$\mu$Jy at $t_{\rm rest} = 2.65$\,yr. We adopt a 6\,GHz upper limit of $<60$\,$\mu$Jy at 1.5\,yr.
\end{enumerate}

There are three events (AT2018hco, AT2019ehz, and AT2019teq) for which the first radio detections occur at $t_{\rm rest} > 2$\,yr. Their earlier non-detections at $t_{\rm rest} < 1$\,yr, combined with declining late-time light curves, prevent reliable interpolation to $\sim1.5$\,yr. To be conservative, we extrapolate their $>2$\,yr declining light curves back to $t_{\rm rest}=1.5$\,yr assuming a power-law decline of the form $f_\nu \propto t^{-\alpha}$ ($\alpha <6$), and treat the result as an upper limit. This gives 6\,GHz upper limits of $<1.2$\,mJy for AT2018hco, $<2.4$\,mJy for AT2019ehz, and $<9.1$\,mJy for AT2019teq. 

\bibliography{main}{}
\bibliographystyle{aasjournalv7}



\end{document}